# Bio-Image Informatics Index BIII: A unique database of image analysis tools and workflows for and by the bioimaging community

Chong Zhang, Alban Gaignard, Matus Kalas, Florian Levet, Felipe Delestro, Joakim Lindblad, Natasa Sladoje,Laure Plantard, Alain Latour, Robert Haase, Gabriel Martins, Paula Sampaio, Leandro Scholz, **NEUBIAS taggers**, Sébastien Tosi, Kota Miura, Julien Colombelli, Perrine Paul-Gilloteaux*

**NEUBIAS taggers: supplementary Table 1**

*** Corresponding author** perrine.paul-gilloteaux@univ-nantes.fr

Bio-image informatics is now commonly defined as the field developing methods and software for processing, analyzing, managing and visualizing biological data from digital imaging systems used in life-science[1–3]. Advanced bio-image informatics methods have now reached the life-science community, mainly thanks to the availability of these methods as software tools. These tools can be available as part of a generic software platform, such as Fiji[4], Icy[5] and Napari[6]. They can also be distributed as independent, stand-alone software, sometimes very specific to a modality of microscopy e.g. super resolution image data[7,8], or specific to a task such as Ilastik[9] offering a panel of machine learning based tools for the segmentation task. Thousands of task specifics tools has been and will be developed either as plugin or standalone software, with diverse way of diffusions: from focused method publication, methods section in a biology focused publication, or directly using a repository hosting and website. The consequence is that finding if an adequate tool exists among the plethora available is tedious. In addition there is a need in the bioimaging community to better document and standardize data description, including their processing[10].

Bio-image analysts, self-defined now thanks to the NEUBIAS initiative, rely on their knowledge of biology, imaging systems and image analysis to analyze bio-imaging data by combining and adapting existing tools into functional workflows to solve particular biological questions[11]. To help the bio-imaging community, a registry was developed by bio-image analysts to provide 1) a unique entry point to search and compare available software for specific task or imaging modality and 2) annotations of each bio-image analysis tool with common vocabularies and definitions related to this field (“tags”).

This registry “Bio Image Informatics Index” BIII (www.biii.eu) is an ongoing and first large crowd-sourced effort to bridge the communities of algorithm and software developers, bio-image analysts and biologists, as a web-based knowledge database (**Figure 1 a-b**). Software tools (for now > 1350), image databases for benchmarking (>25) and training materials (>78) for bio-image analysis are referenced and curated following standardized vocabularies.

The curation is conducted by “Taggers” i.e. biologists, bio-image analysts, microscopists and software developers editing the database (**supplementary 1**). The curation of the information, the development of the web tool itself and the definition of standards was mostly undergone during unique events called “Taggathons”. Similar to “Hackathons” where developers gather to create or improve existing code, “taggers” gather during Taggathons to "tag" the entries in BIII and help developing and organizing this online repository. Outside Taggathons, NEUBIAS training events participants[12] were also invited to follow guidelines to add their own tools or their favorite software description to become “taggers”.

When “tagging”, a tagger first defines the type of the software tool under consideration: this can either be a workflow (e.g. a script for addressing a specific biological question), a workflow component (e.g. an implementation of certain image processing or analysis algorithm), or a component collection (e.g. a library, with or without a platform)[11]. Next, the tagger adds “tags” which aim to enhance discoverability using two ontologies (**supplementary 2**). The core ontology defines the fields describing a tool. These fields include technical information such as the computer language used and its dependencies, information about the authors, a short usage description of the tool, the download URL, how to cite it, and additional links such as related training material. The main imaging modalities of applications, or image processing operations performed by this tool are also described using controlled vocabularies. The controlled vocabularies are selected within Edam Bio Imaging[13], a unique ontology for bio-image analysis which was created for BIII during the Taggathons and was defined iteratively by examining proposed new tags in BIII. Finally, “workflow” entries can be described as a flowchart, with referencing to other existing entries in BIII (e.g. components). Indeed, one of the purposes of BIII is to help the findability of tools in particular with the idea of combining them to achieve a bio image analysis workflow for a specific question.

One of the prominent difficulties we observed in interdisciplinary fields such as bio-image analysis is the large gap in the vocabularies and concepts used in biological and computational fields. Due to this gap, the regular web search for resources tends to become inefficient. BIII aims to solve this problem and help closing the vocabulary gaps from biology to image processing. For example “centromere”, a biological term not commonly known by image processing experts, can be linked within BIII database to “Spot Detection”, image processing term, or “microtubule” (biological term) to “Filament Tracing” (image processing term). Furthermore, such cross-disciplinary linking allows discovering and applying tools with similar aims to other biological objects: a tool originally developed for blood vessels or neuron dendrites segmentation could reveal to be useful for segmenting microtubules as well because the image processing operation is similar (**Figure 1 c d**).

BIII also helps to avoid duplicating efforts in creating database for bio-image analysis software. For example, similar initiatives already exist for a broader range of applications[14], and for more specific topics such as plant analysis[15]. For reusability, all entries are exposed by following FAIR principles in order to allow them to be accessed for other usage under an open data license. This is for example in use in the correlation-software list established by the Correlated Multimodal Imaging in Life Science network COMULIS. Further list can be created and exported using a gathering term such as “Single molecule localization microscopy” (**supplementary 3**). We considered that each community can have a different interest and that data can be exposed in different ways by sub selecting the information of interest. Thanks to this data exposition, the search in BIII can also be conveniently performed directly from NEUBIAS Fiji Search bar, available from the FIJI NEUBIAS update site, and could be extended to other software platform.

Another added value of the exposition of data from BIII is to enable the data mining of the bio-image analysis field itself. This could for instance allow highlighting missing software interoperability, revealing image analysis algorithms yet to be used in the biological field, or finding out most common tasks associated with imaging modalities. Some preliminary examples of data mining are provided in **supplementary 3**.

Finally, BIII provides a list of training resource materials in bio image analysis for self-education or for trainers as well as pointer to image data sets to test tools.

The long-term success and growth of BIII will be greatly enhanced by further communities’ involvement. The planned roadmap to extend BIII functionalities will support the consolidation of bioimage analysis as a high-yielding research field for Life Science. We call the community to contribute online by curating content or organizing Taggathons event, for example as satellite events in communities gathering.

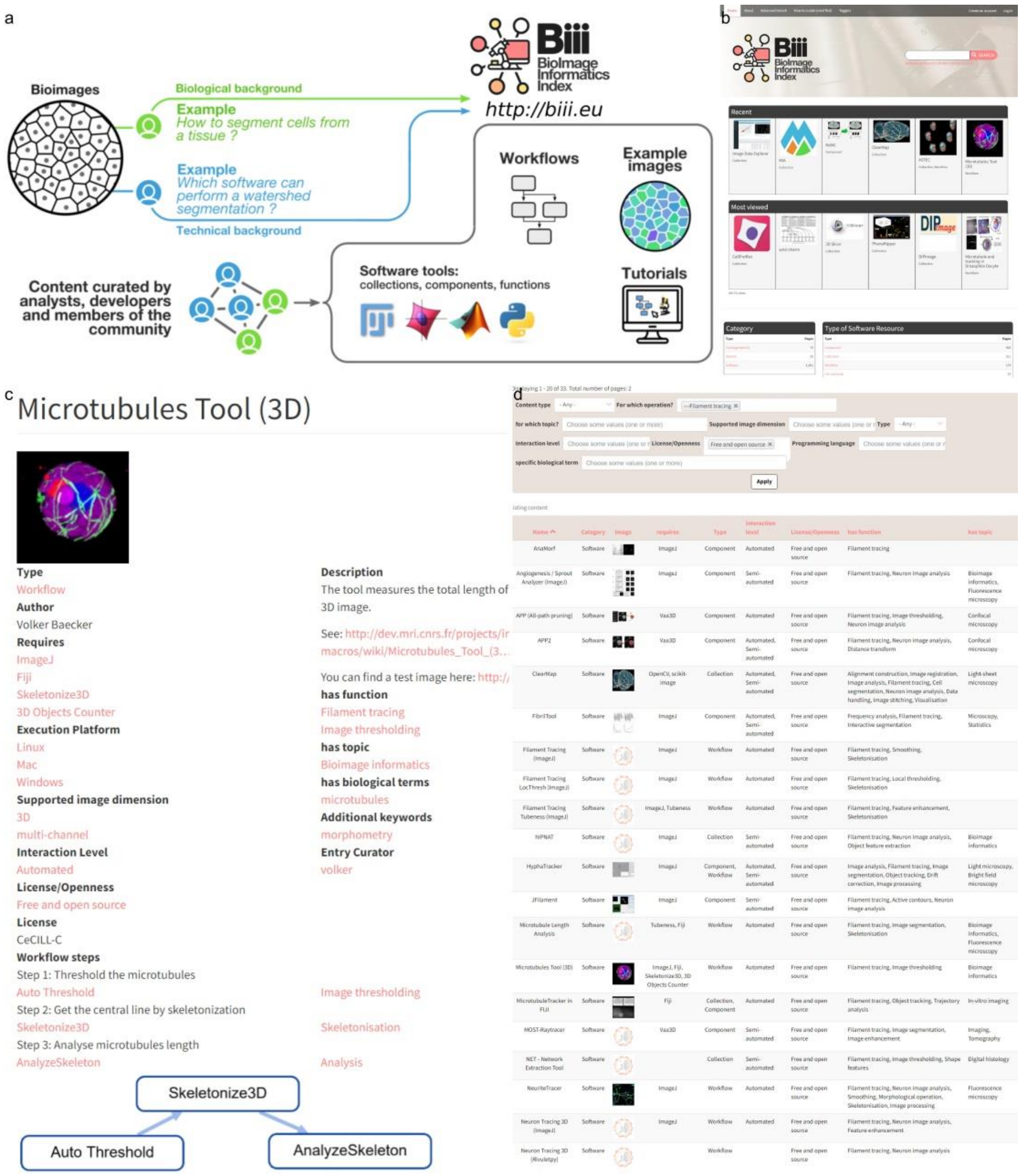

*Figure 1:* ***a)*** *general concept of BIII.* ***b)*** *Home page of BIII, showing the most recently modified or added entries in the database and the most viewed, as well as the total number of entries by categories.* ***c)*** *Example software tool description, here a workflow for microtubule length analysis Microtubules Tool (3D). The interactive graphical representation is generated from the steps of the workflow as entered by the curator.* ***d)*** *The first results of advanced search page in BIII for Operation « Filament Tracing », restrained to the " Free and Open Source software" level of openness showing result for neuron as well as microtubules analysis.*

## Acknowledgements

Most taggathons were funded by COST CA15124 (NEUBIAS). The taggathons for the COMULIS bridge was funded by COST CA17121. The authors also acknowledge the support of the France-BioImaging infrastructure (ANR-10-INBS-04), l'institut Francais de Bio Informatique (ANR-11-INBS-0013) , IRB

Barcelona, the MicroPICell facility (Ibisa,Biogenouest) member of France BioImaging, Eurobioimaging We are grateful to ELIXIR to have recognized BIII as a Recommended Interoperability Resources since 2021. Local organizers of the taggathons event who have fostered the development of BIII are also greatly acknowledged: Jean Salamero, Julia Fernandez, Peter Horvath, Bertrand Vernay, Aymeric Fouquier d'Herouel, Andreas Girod, Fabrice Cordelières. RH acknowledges support by the Deutsche Forschungsgemeinschaft under Germany's Excellence Strategy – EXC2068 - Cluster of Excellence Physics of Life of TU Dresden. A research scholarship was granted to Leandro Scholz by CAPES (Coordenação de Aperfeiçoamento de Pessoal de Nível Superior), a Brazilian government agency for the development of personnel in higher education.

## Data availability

This database content of BIII is made available under the Open Data Commons Attribution License (ODC-By) v1.0 https://opendatacommons.org/licenses/by/1.0/.

## Code availability

The implementation of the BIII web platform is based on the CMS Drupal 8, with some additional custom modules. The full source code and collaborative development environment of the website is available on GIT HUB https://github.com/NeuBIAS/bise/wiki

Code for the NEUBIAS search bar is available https://github.com/imagej/imagej-plugins-search-biii

Code for demoing advanced request based on Jupyter notebook is available NEUBIAS/bise-core-ontology (github.com)

## Supplementary Information

**Supplementary Table 1:** List of NEUBIAS taggers.

**Supplementary Note 2:** Description of the two ontologies NEUBIAS Core Ontology and Edam Bio Imaging.

**Supplementary Note 3:** Examples of advanced search and of advanced queries exploiting BIII content for data mining.

| Surname | Name | Gender | Country | Affiliation |
|---|---|---|---|---|
| Baecker | Volker | M | France | MRI - CNRS |
| Ball | Graeme | M | UK | University od Dundee |
| Bankhead | Peter | M | UK | University of Edinburgh |
| Barry | Joseph | M | Germany | EMBL - Heidelberg |
| Beard | Niall | M | UK | University of Manchester |
| Berg | Stuart | M | USA | HHMI Janelia |
| Bianchini | Paolo | M | Italy | Italian Institute of Technology |
| Boizeau | Marie-Laure | F | France | ITAV Toulouse |
| Burri | Oliver | M | Switzerland | EPFL - Lausanne |
| Caldas | Victor | M | Netherlands | AMOLF - Vrije Universiteit Amsterdam |
| Cardone | Giovanni | M | Germany | Max Planck Institute Biochemistry, Munich |
| Chessel | Anatole | M | France | École Polytechnique - Palaiseau |
| Cinquin | Bertrand | M | France | École Normale Supérieure - Cachan |
| Colombelli | Julien | M | Spain | IRB Barcelona |
| Cordelières | Fabrice | M | France | IINS - University of Bordeaux |
| de Chaumont | Fabrice | M | France | Pasteur Institute - Paris |
| Delestro | Felipe | M | France | École Normale Supérieure |
| Delgado | Ricard | M | Switzerland | EPFL - Lausanne |
| Dominguez Del Angel | Victoria | F | France | Institut Français de Bioinformatique |
| Dufour | Alexandre | M | France | Institut Pasteur Paris |
| Eglinger | Jan | M | Switzerland | FMI - Basel |
| Fondón | Irene | F | Spain | University of Seville |
| Gaboriau | David | M | UK | Imperial College - London |
| Gaignard | Alban | M | France | University of Nantes |
| Gelman | Laurent | M | Switzerland | FMI, Basel |
| Golani | Ofra | F | Israel | Weizmann Institute - Rehovot |
| Guiet | Romain | M | Switzerland | EPFL - Lausanne |
| Haase | Robert | M | Germany | TU Dresden |
| Horn | Martin | M | Germany | University of Konstanz |
| Ison | Jon | M | Denmark | Technical University of Denmark |
| Jones | Martin | M | UK | Francis Crick Institute - London |
| Kain | Mickael | M | France | IRISA Rennes |
| Kalas | Matus | M | Norway | University of Bergen |
| Kamentsky | Lee | M | USA | Broad Institute - Cambridge, MA, USA |
| Kankaanpää | Pasi | M | Finland | University of Turku |
| Kirschmann | Moritz | M | Switzerland | ZMB University of Zuerich |
| Kladt | Nikolay | M | Germany | Univeristy of Koeln |
| Koethe | Ullrich | M | Germany | Univeristy of Koeln |
| Latour | Alain | M | France | Université de Lorraine |
| Legland | David | M | France | BIBS INRAE Nantes |
| Leng | Joanna | F | UK | University of Leeds |
| Levet | Florian | M | France | IINS Bordeaux |
| Ligteringen | Ronald | M | Netherlands | TU Delft |
| Lindblad | Joakim | M | Sweden | University of Uppsala |
| Lobet | Guillaume | M | Germany | Forschungszentrum Juelich |
| Loo | Kevin | M | The Czech Republic | Czech Technical University. |
| Majer | Peter | M | Switzerland | Bitplane AG (Oxford Instruments) |
| Marée | Raphael | M | Belgium | Université de Liège |
| Martins | Gabriel | M | Portugal | IGC - Oeiras |
| Maulucci | Giuseppe | M | Italy | Università cattolica del Sacro Cuore - Rome |
| Maumet | Camille | F | France | INRIA Rennes |
| Ménager | Hervé | M | France | Institut Pasteur Paris |
| Miura | Kota | M | Germany | University of Heidelberg |
| Moehl | Christoph | M | Germany | DZNE - Bonn |
| Moore | Josh | M | UK | University of Dundee |
| Norrelykke | Simon | M | Switzerland | ETH Zuerich |
| Ortiz de Solórzano Aurusa | Carlos | M | Spain | CIMA - Pamplona |
| Ouyang | Wei | M | Sweden | KTH Royal Institute of Technology |
| Paul-Gilloteaux | Perrine | F | France | University of Nantes |
| Pavie | Benjamin | M | Belgium | VIB Leuven |
| Pavolainen | Lassi | M | Finland | FIMM - Helsinki |
| Pedro Coelho | Luis | M | Germany | EMBL Heidelberg |
| Pengo | Thomas | M | USA | Univeristy of Minnesota |
| Pike | Jeremy | M | UK | CRUK Cambridge |
| Plantard | Laure | F | Denmark | FMI Basel |
| Ponti | Aaron | M | Switzerland | ETHZ BSSE Basel |
| Prigent | Sylvain | M | France | University of Rennes |
| Ranefall | Petter | M | Sweden | SciLifeLab Stockholm |
| Rey-Villamizar | Nicolas | M | USA | University of Houston |
| Rousseau | David | M | France | University of Angers |
| Rueden | Curtis | M | USA | University of Madison |
| Rusticci | Gabriella | F | UK | EMBL-EBI |
| Sage | Daniel | M | Switzerland | EPFL Lausanne |
| Sampaio | Paula | F | Portugal | IBMC Porto |
| Sarmiento | Auxiliadora | F | Spain | University of Seville |
| Sheehan Rooney | Kelly | F | Germany | Eurobioimaging |
| Schindelin | Johannes | M | USA | University of Madison |
| Scholz | Leandro Aluisic | M | Brasil | University of Paraná |
| Schulze | Keith | M | Australia | Monash University |
| Signolle | Nicolas | M | France | Institut de Cancerologie Gustave Roussy |
| Sladoje | Natasa | F | Serbia | University of Uppsala |
| Sommer | Christoph | M | Germany | IMBA Wien |
| Stoeter | Torsten | M | Germany | LIN Magdeburg |
| Stoma | Szymon | M | Switzerland | ETH Zuerich |
| Swedlow | Jason | M | UK | University od Dundee |
| Thierry | Raphael | M | Switzerland | FMI Basel |
| Tischer | Christian | M | Germany | EMBL Heidelberg |
| Tosi | Sébastien | M | Spain | IRB Barcelona |
| Marak | Laszlo | M | France | ujoimro@gmail.com |
| Unay | Devrim | M | Turkey | İzmir Demokrasi University |
| Waithe | Dominic | M | UK | University College London |
| Walczysko | Petr | M | UK | University of Dundee |
| Walter | Thomas | M | France | Mines Paristech |
| Weber | Igor | M | Croatia | IRB - Zagreb |
| Yagüe | Carlos | M | Spain | Universitat Pompeu Fabra - Barcelona |
| Zhang | Chong | F | Spain | Universitat Pompeu Fabra - Barcelona |

# Description of the two ontologies in use in BIII:

## NEUBIAS Core ontology

NEUBIAS core ontology standardizes the fields used to describe a software tool in image analysis. Some of the fields are generic for a software tool, such as execution platform, license, download page or reference publications. Others are very dependent of the bio-image analysis field, such as the supported image dimension (2D, 3D, multi-channel, time) or the topic and operation to be selected from the EDAM bio-imaging ontology. They are organized in five main sections: main information such as its description and authors; Links information gathering download, usage, or related training material link; Tags which aims to enhance the tools discoverability and consists of using a controlled vocabulary from edam bio imaging ontologies; Usage which gives technical information such as the language used or its dependencies; Workflow types and descriptions of each workflow step referring to other existing entries.

The documentation is available here: https://github.com/NEUBIAS/bise-core-ontology

Recommandation for completion of curation are described in the How to Curate guidelines http://biii.eu/howtocurate as shown in figure S1 and S2.

A graphical visualization can be generated and parsed using VOWL[1]:

http://vowl.visualdataweb.org/webvowl-old/webvowl-old.html#iri=https://raw.githubusercontent.com/NeuBIAS/bise-core-ontology/master/owl-ontology/bise-core-ontology-v1.1.owl (Figure S3).

| SOFTWARE Entry | | | |
|---|---|---|---|
| **Information** | **Sparse** | **Detailed** | **Comprehensive** |
| **Name** | x | x | x |
| **Description** | x | x | x |
| **Unique ID** | x | x | x |
| **Type (1)** | x | x | x |
| **has implementation type (2)** | | x | x |
| **Entry point (3)** | | x | x |
| **has illustrative image** | | x | x |
| **has author [LastName, FirstName (ORCID)]** | | x | x |
| ***has function (EDAM-Bioimaging Operation)**** | | x | x |
| ***Accessibility (see groups)*** | | x | x |
| ***Use information (see groups)*** | | x | x |
| ***has programming language*** | | x | x |
| **execution platform (operating system)** | | x | x |
| **has topic (EDAM-Bioimaging Topic)*** | | x | x |
| ***Documentation (see groups)*** | | | x |
| **has DOI (4)**** | | | x |
| **Training Material (see groups)** | | | x |
| ***has biological terms**** | | | not mandatory |
| **Additional Keywords*** | | | not mandatory |

***** At least one  ** Still uncommon (optional)

(1) (Collection, Component, Workflow, Don't know)

(2) Options: Library, Plugin, Standalone and Web

(3) Repository website or other 'official' website

(4)  DOI of the implementation, not the publication

***Figure S1:*** *Recommendations regarding the completeness of software entries based on the core ontology.*

**Groups**

| | |
|---|---|
| **Accessibility** | |
| has license (text field)<br>license / openness | *At least one* |
| **Use information** | |
| Supported image dimension<br>requires | *At least one* |
| **Documentation** | |
| has Reference Publication Link<br>has Documentation Link<br>has Comparison | *At least one* |
| **Training Material** | |
| has Usage Example Link<br>Training material (in biii.eu) | *At least one* |

***Figure S2**: Definition of groups in the core ontology.*

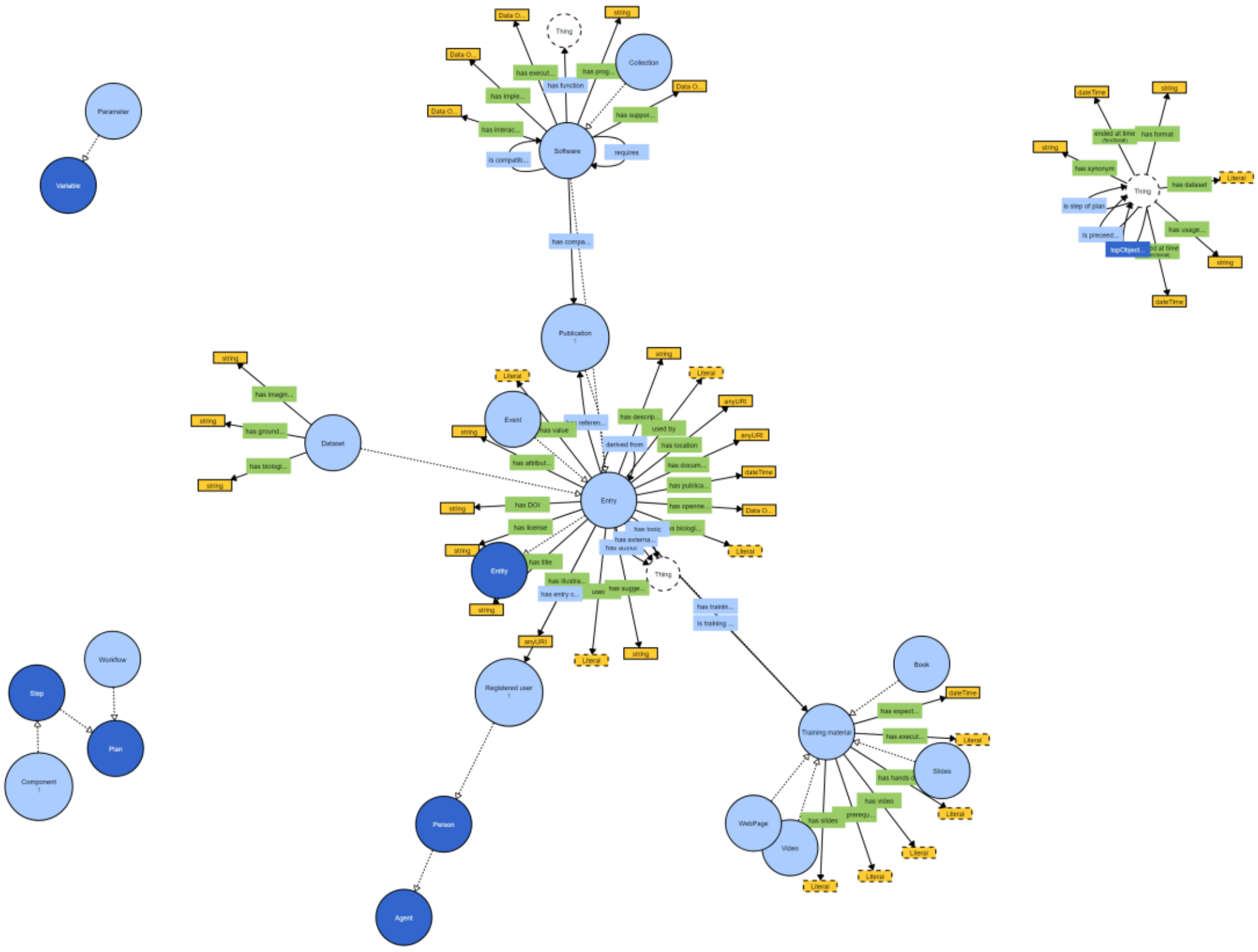

***Figure S3:*** *Automatic documentation on the BIII core ontology using graphical representation as allowed by Webvowl.*

## EDAM-Bioimaging

EDAM-bioimaging[2–4] is an extension of the EDAM ontology[5] , dedicated to bio-image analysis, bio-image informatics, and bio-imaging. It then follows the same main concepts of Topics, Operation, Data and Format (**Figure S4**) to be compatible with EDAM. EDAM-bioimaging contains then an inter-related hierarchy of concepts including bio-image analysis and related operations, bio-imaging topics and technologies, and bioimage data and their formats.

The ontology can be parsed from Bioportal[6] (**Figure S5**): https://bioportal.bioontology.org/ontologies/EDAM-BIOIMAGING/?p=classes&conceptid=http%3A%2F%2Fedamontology.org%2Foperation_Image_deconvolution#visualization.

**Figure S6** shows two examples of topic and operation concepts. The modelled concepts enable interoperable descriptions of software, publications, data, and workflows, fostering reliable, transparent and "reproducible" bio-image analysis.

EDAM-bioimaging is under active development, with a couple of alpha releases publicly available, as for instance alpha06 (**Figure S7**). It is developed by a large crowd-sourced effort in a welcoming collaboration between bio-imaging experts and ontology developers. It is used in BIII to describe the applications of these tools, by describing the operations performed (such as segmentation, visualization, or lower level operation) and the field of applications of these tools such as the imaging modalities to which it can be applied.

The terms are defined and discussed during specific taggathons, with bio-imaging experts and ontology developers. It can be contributed by adding suggestions of terms directly in BIII using the tag category “ADDITIONAL KEYWORDS” if the bio edam ontology tags were not sufficient. They will be considered for extending the bio edam ontology.

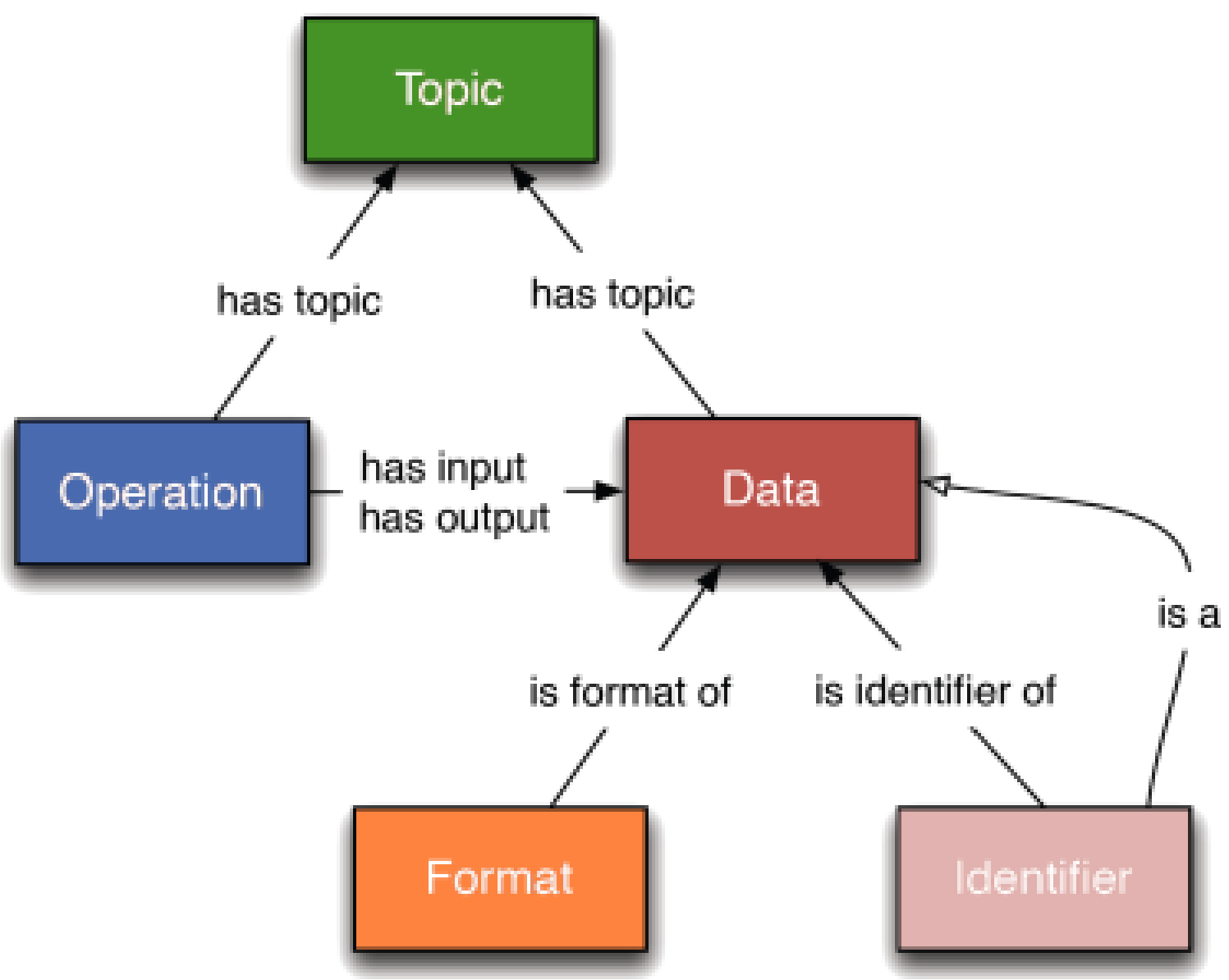

***Figure S4*** *: Structure of EDAM[5]*

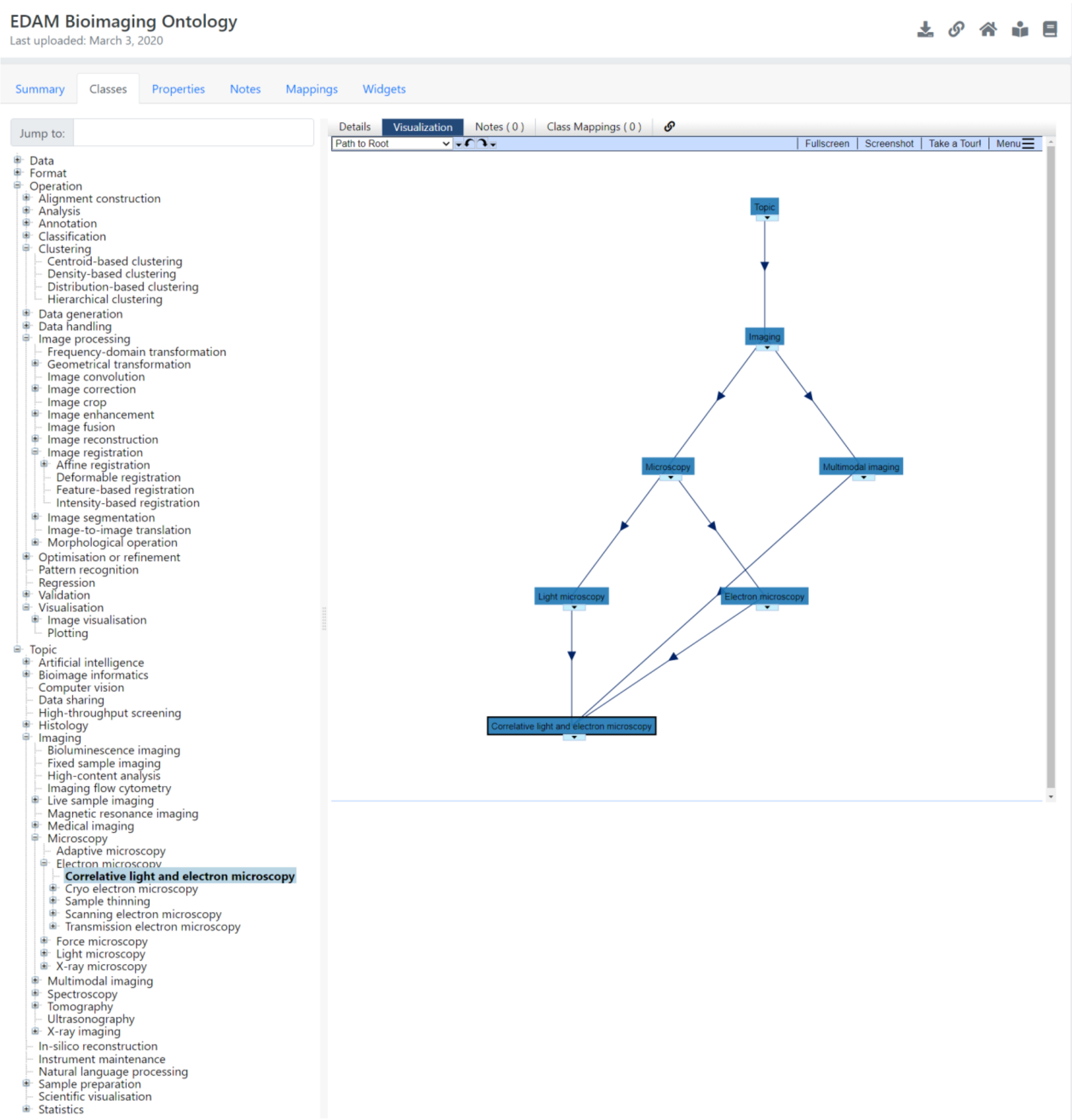

***Figure S5:*** *Example of access to the EDAM BIOIMAGING ontology using BioPortal*[6]

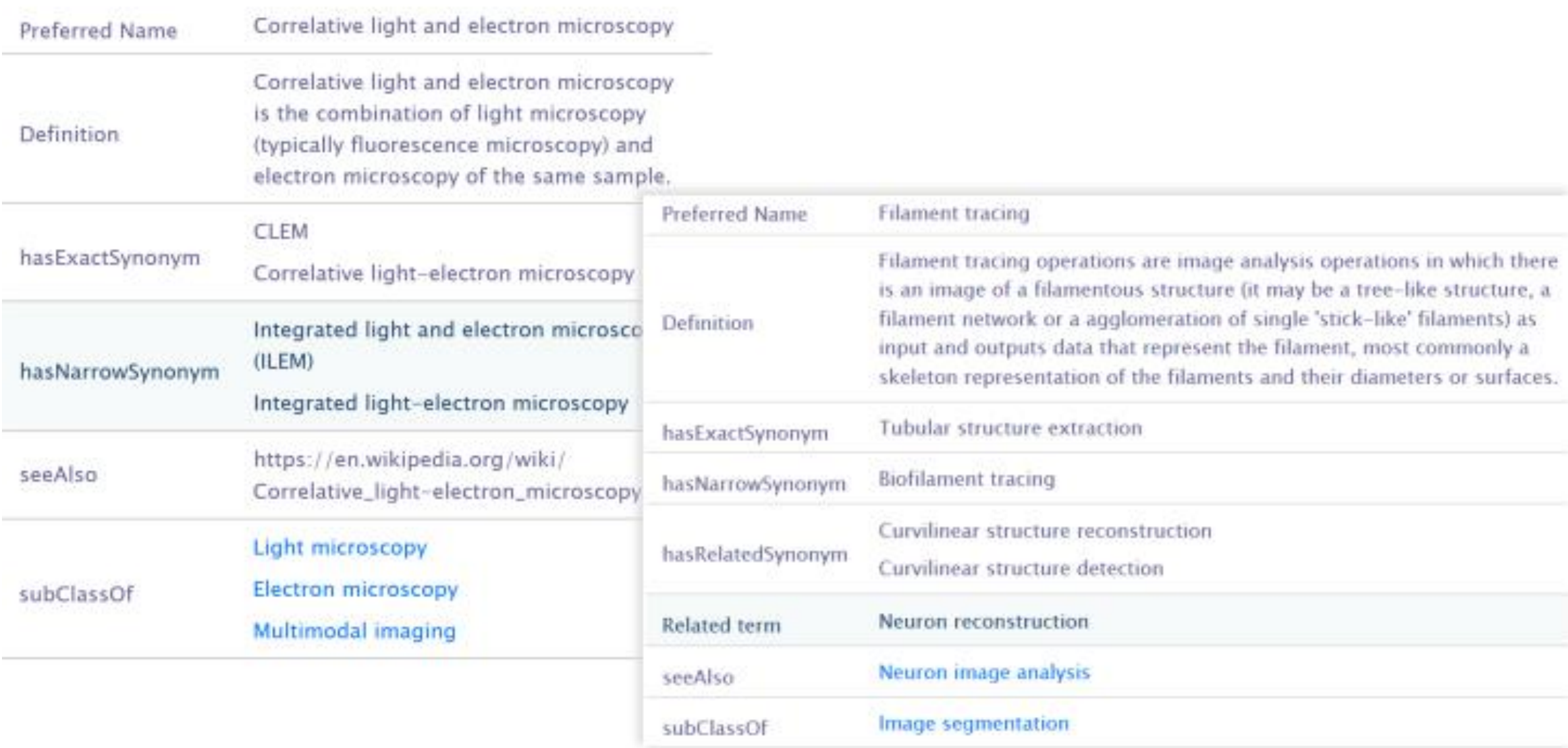

| | |
|---|---|
| Preferred Name | Correlative light and electron microscopy |
| Definition | Correlative light and electron microscopy is the combination of light microscopy (typically fluorescence microscopy) and electron microscopy of the same sample. |
| hasExactSynonym | CLEM<br>Correlative light-electron microscopy |
| hasNarrowSynonym | Integrated light and electron microsco (ILEM)<br>Integrated light-electron microscopy |
| seeAlso | https://en.wikipedia.org/wiki/Correlative_light-electron_microscopy |
| subClassOf | Light microscopy<br>Electron microscopy<br>Multimodal imaging |
| Preferred Name | Filament tracing |
| Definition | Filament tracing operations are image analysis operations in which there is an image of a filamentous structure (it may be a tree-like structure, a filament network or a agglomeration of single 'stick-like' filaments) as input and outputs data that represent the filament, most commonly a skeleton representation of the filaments and their diameters or surfaces. |
| hasExactSynonym | Tubular structure extraction |
| hasNarrowSynonym | Biofilament tracing |
| hasRelatedSynonym | Curvilinear structure reconstruction<br>Curvilinear structure detection |
| Related term | Neuron reconstruction |
| seeAlso | Neuron image analysis |
| subClassOf | Image segmentation |

***Figure S6:*** *Two example concepts from EDAM-BioImaging: "Correlative Light And Electron Microscopy" topic and "Filament Tracing" Operation.*

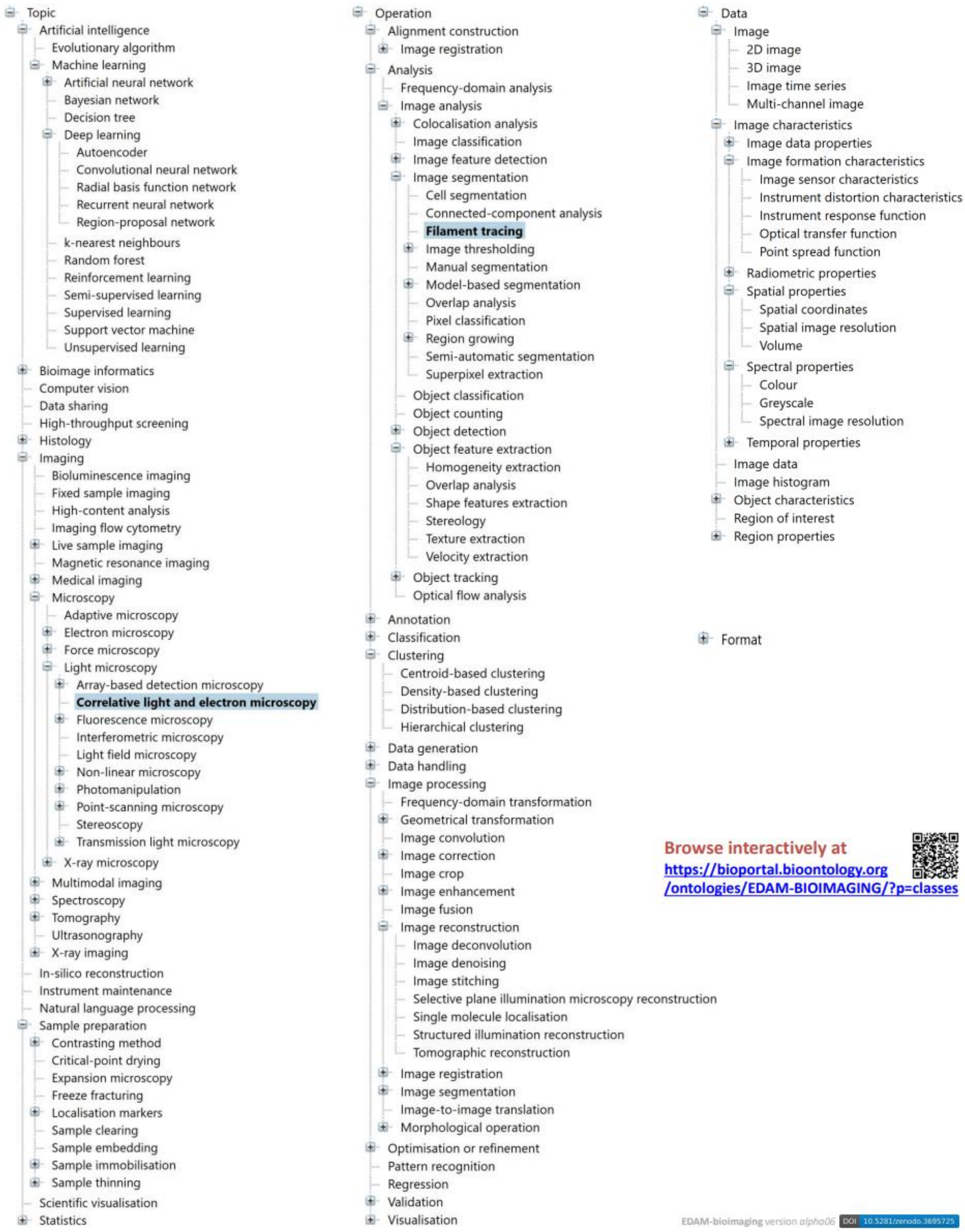

***Figure S7:*** *Hierarchies in EDAM-Bioimaging as in alpha06 version.*

# Examples of advanced search and of advanced queries exploiting BIII content for data mining.

# 1.Using Advanced Search

## 1.1 Example to generate list of software for a specific modality of microscopy

When looking for a tool, the advanced search page all content | BIII https://biii.eu/all-content allows to filter entries based on the ontology in use (Supplementary Note 2 “Description of the two ontologies in use in BIII”). For instance, one could list all the entries related to the topic “Single Molecule Localization Microscopy” to see at a glance the list of entries tagged with this term (Figure S1). The term is also defined https://biii.eu/single-molecule-localization-microscopy by importing EDAM-Bioimaging ontology, including synonyms. The usage of synonyms is not implemented yet in the current version of BIII but this implementation is part of the development roadmap. Note that the localization task itself is an operation https://biii.eu/single-molecule-localisation .

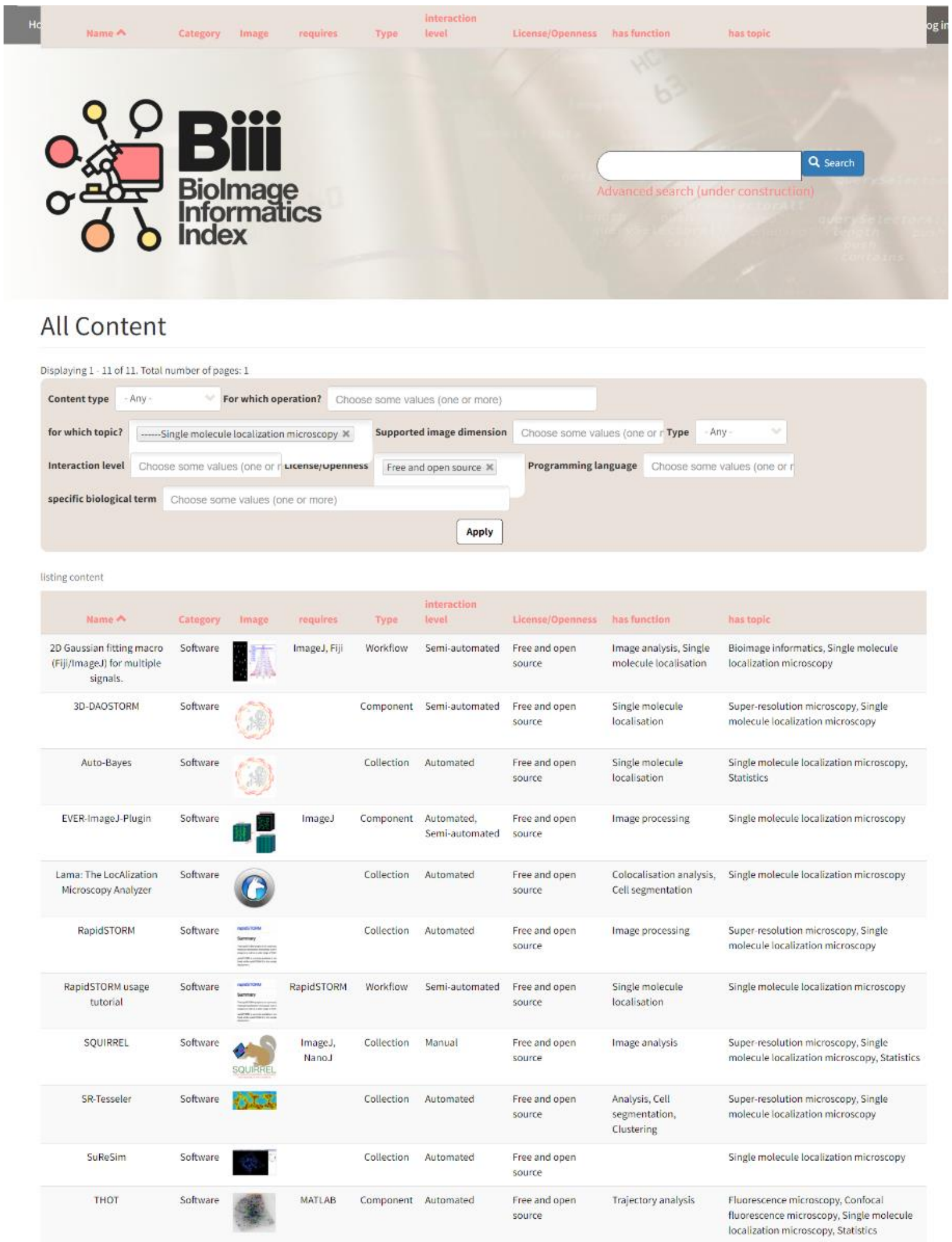

***Figure S1:*** *Result of search performed on the topic Single Molecule Localization Microscopy, sending a list of software tools, data set and training material related to this topic, i.e. having been developed or used for this particular microscopy modality. Here the search was restricted to Free and Open source tools and truncated for display purpose, the full list is available at* https://biii.eu/all-content?type=All&field_has_topic_target_id_1%5B%5D=4102&field_type_target_id=All&field_license_openness_target_id%5B%5D=3575

## 1.2 Looking for software: example for benchmarking under BIAFLOWS

BIAFLOWS[1] is a benchmarking platform which was developed under the auspices of NEUBIAS.

To enhance their visibility, all workflows hosted in the system are also referenced from BIII, and the mention of comparison exists as the “has comparison” field.

For instance, looking in the biii search bar for Biaflows&Nuclei&segmentation will return a list of entries (Figure S2) which has been benchmarked in BIAFLOWS for nuclei segmentation (Figure S3)

Note that the space bar denotes “+” search meaning OR and will return more results as it will list all tools containing at least one of the listed terms. Adding “&” in the search means AND, and will return tools matching the list of terms.

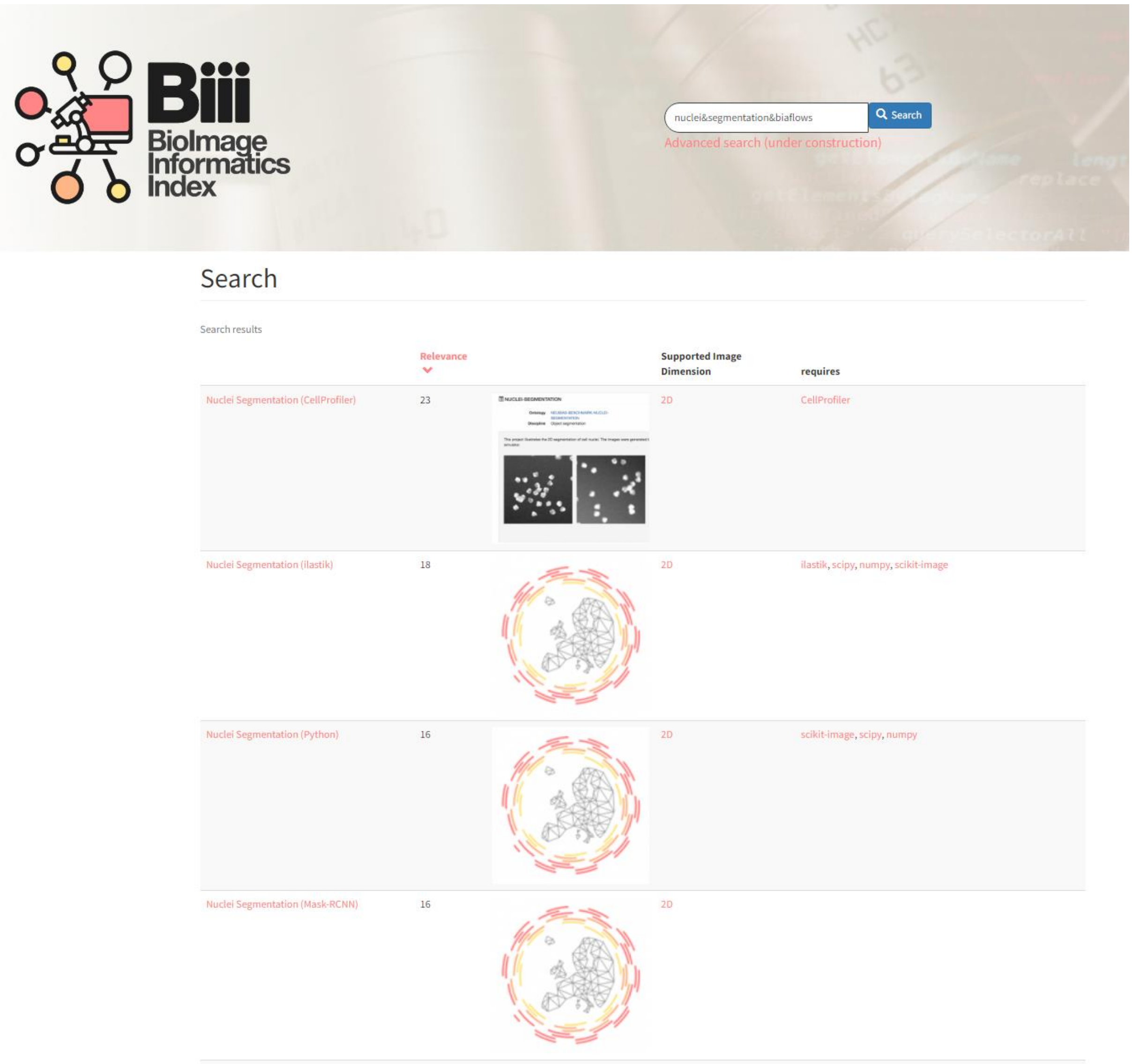

***Figure S2:*** *Search return looking for nuclei&segmentation&biaflows. The list has been truncated for display purpose, the full list is accessible from [http://biii.eu/search?search_api_fulltext=biaflows%26nuclei%26segmentation](http://biii.eu/search?search_api_fulltext=biaflows%26nuclei%26segmentation).*

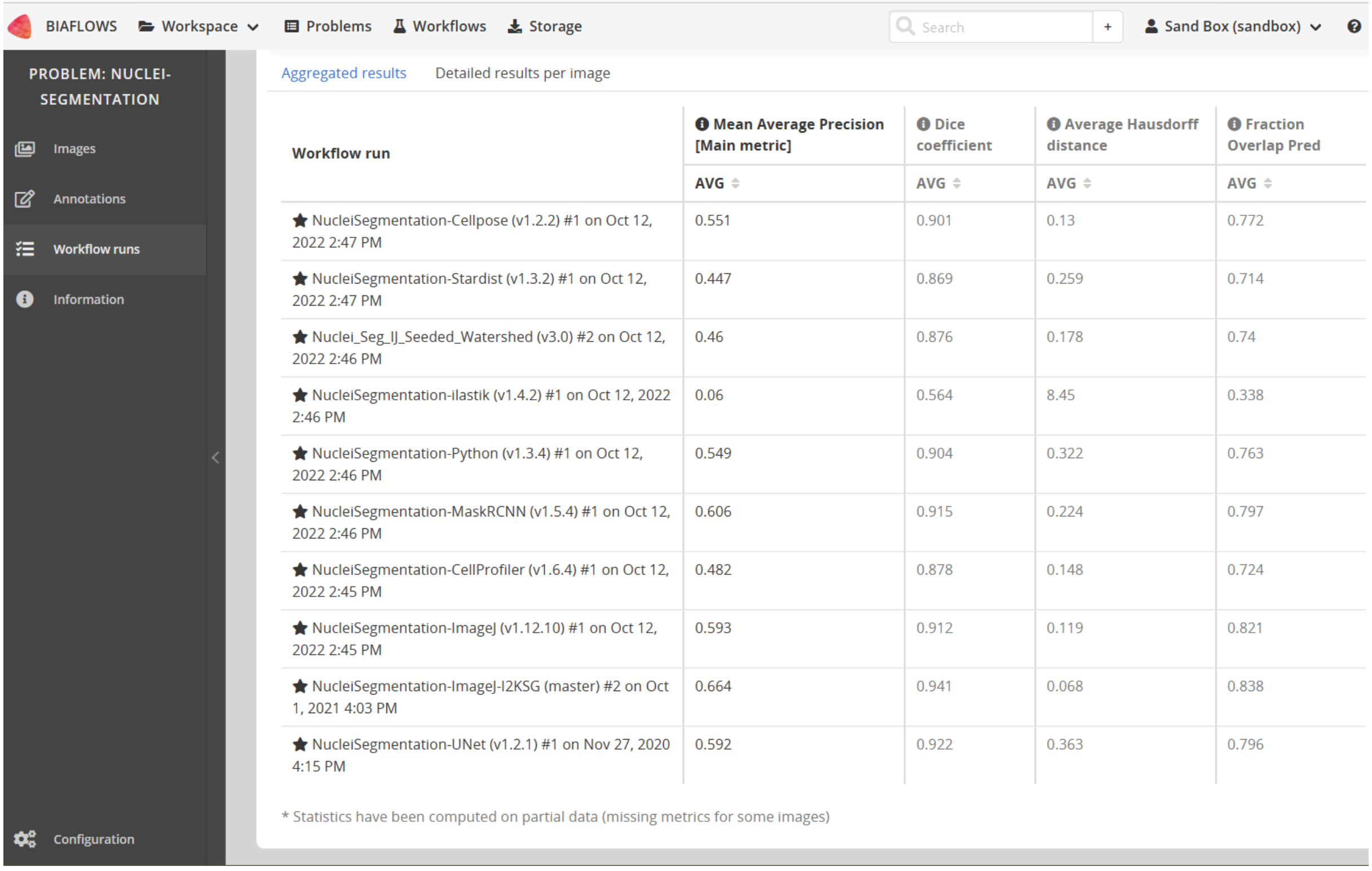

| Workflow run | Mean Average Precision [Main metric] AVG | Dice coefficient AVG | Average Hausdorff distance AVG | Fraction Overlap Pred AVG |
|---|---|---|---|---|
| NucleiSegmentation-Cellpose (v1.2.2) #1 on Oct 12, 2022 2:47 PM | 0.551 | 0.901 | 0.13 | 0.772 |
| NucleiSegmentation-Stardist (v1.3.2) #1 on Oct 12, 2022 2:47 PM | 0.447 | 0.869 | 0.259 | 0.714 |
| Nuclei_Seg_IJ_Seeded_Watershed (v3.0) #2 on Oct 12, 2022 2:46 PM | 0.46 | 0.876 | 0.178 | 0.74 |
| NucleiSegmentation-ilastik (v1.4.2) #1 on Oct 12, 2022 2:46 PM | 0.06 | 0.564 | 8.45 | 0.338 |
| NucleiSegmentation-Python (v1.3.4) #1 on Oct 12, 2022 2:46 PM | 0.549 | 0.904 | 0.322 | 0.763 |
| NucleiSegmentation-MaskRCNN (v1.5.4) #1 on Oct 12, 2022 2:46 PM | 0.606 | 0.915 | 0.224 | 0.797 |
| NucleiSegmentation-CellProfiler (v1.6.4) #1 on Oct 12, 2022 2:45 PM | 0.482 | 0.878 | 0.148 | 0.724 |
| NucleiSegmentation-ImageJ (v1.12.10) #1 on Oct 12, 2022 2:45 PM | 0.593 | 0.912 | 0.119 | 0.821 |
| NucleiSegmentation-ImageJ-I2KSG (master) #2 on Oct 1, 2021 4:03 PM | 0.664 | 0.941 | 0.068 | 0.838 |
| NucleiSegmentation-UNet (v1.2.1) #1 on Nov 27, 2020 4:15 PM | 0.592 | 0.922 | 0.363 | 0.796 |

* Statistics have been computed on partial data (missing metrics for some images)

***FigureS3:*** *BIAFLOWS benchmarking board for the nuclei segmentation problem accessible from [Biaflows (neubias.org)](https://biaflows-sandbox.neubias.org/#/project/205081/analysis) [https://biaflows-sandbox.neubias.org/#/project/205081/analysis](https://biaflows-sandbox.neubias.org/#/project/205081/analysis) . Workflows are documented in BIII.*

# 2. Example of queries

## 2.1 Using JSON exposure of data

JSON is a standard format to facilitate exchange of data within the web[2]. BIII offers the export of all its data under this format for further usage. We exemplify two of them below.

### 2.1.1 Displaying customized list of software on another webpage

COMULIS (Correlated Multimodal Imaging in Life Sciences) is an EU-funded COST Action that aims at fueling urgently needed collaborations in the field of correlated multimodal imaging (CMI), promoting and disseminating its benefits through showcase pipelines, and paving the way for its technological advancement and implementation as a versatile tool in biological and preclinical research[3]. One of the effort was to list software in use in the multimodal imaging community.

For this purpose, a survey was conducted among the members of the network part of the dedicated working group (92 members over the 305 COMULIS members). All listed tools by these members were then curated in BIII by adding to their description the “Multimodal Imaging” tag for the “has Topic” category. A webpage was created https://www.comulis.eu/correlation-software on the COMULIS website (Figure S4), with three buttons corresponding to the three main operations identified by the member of the network (Segmentation, Registration, Visualization). Clicking on one of this button will ask BIII to return the list of software matching both the topic “Multimodal Imaging “ and the Operation selected, using the JSON exposure of all data , as for instance for segmentation : biii.eu/searchjsonexport?search_api_fulltext=(?=segmentation)&_format=json combined with the Multimodal Imaging (Figure S4). Results of the previous query not matching the multimodal imaging topic are then proposed as “You might also like”.

The source code to generate this page is available from https://github.com/delestro/comulis for reuse.

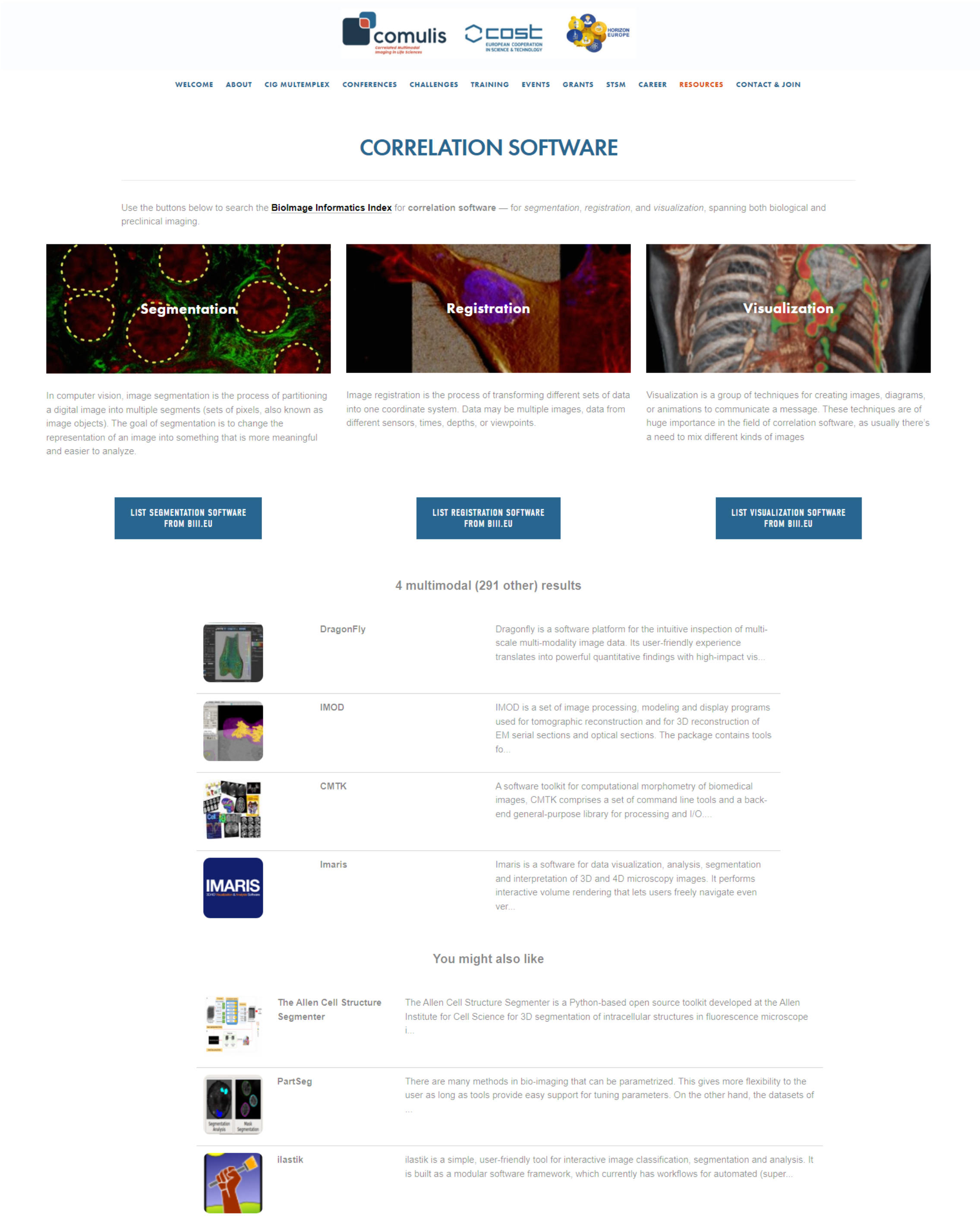

*Figure S4: Screen capture of the https://www.comulis.eu/correlation-software webpage, allowing to keep an updated page of sub selected software based on the curation in BIII.*

### 2.1.2 constructing a search bar in Fiji

Thanks to this data exposition, the search in BIII can also be conveniently performed directly from NEUBIAS Fiji Search bar (Figure S5), available from the FIJI NEUBIAS update site https://sites.imagej.net/Neubias/, and the same concept could be extended to other software platform. In short, it performs a search in BIII restraining the results to the ImageJ platform. This allow for a richer search than the default one because workflows can also be returned as results.

The source code and documentation is available from [https://github.com/imagej/imagej-plugins-search-biii](https://github.com/imagej/imagej-plugins-search-biii)

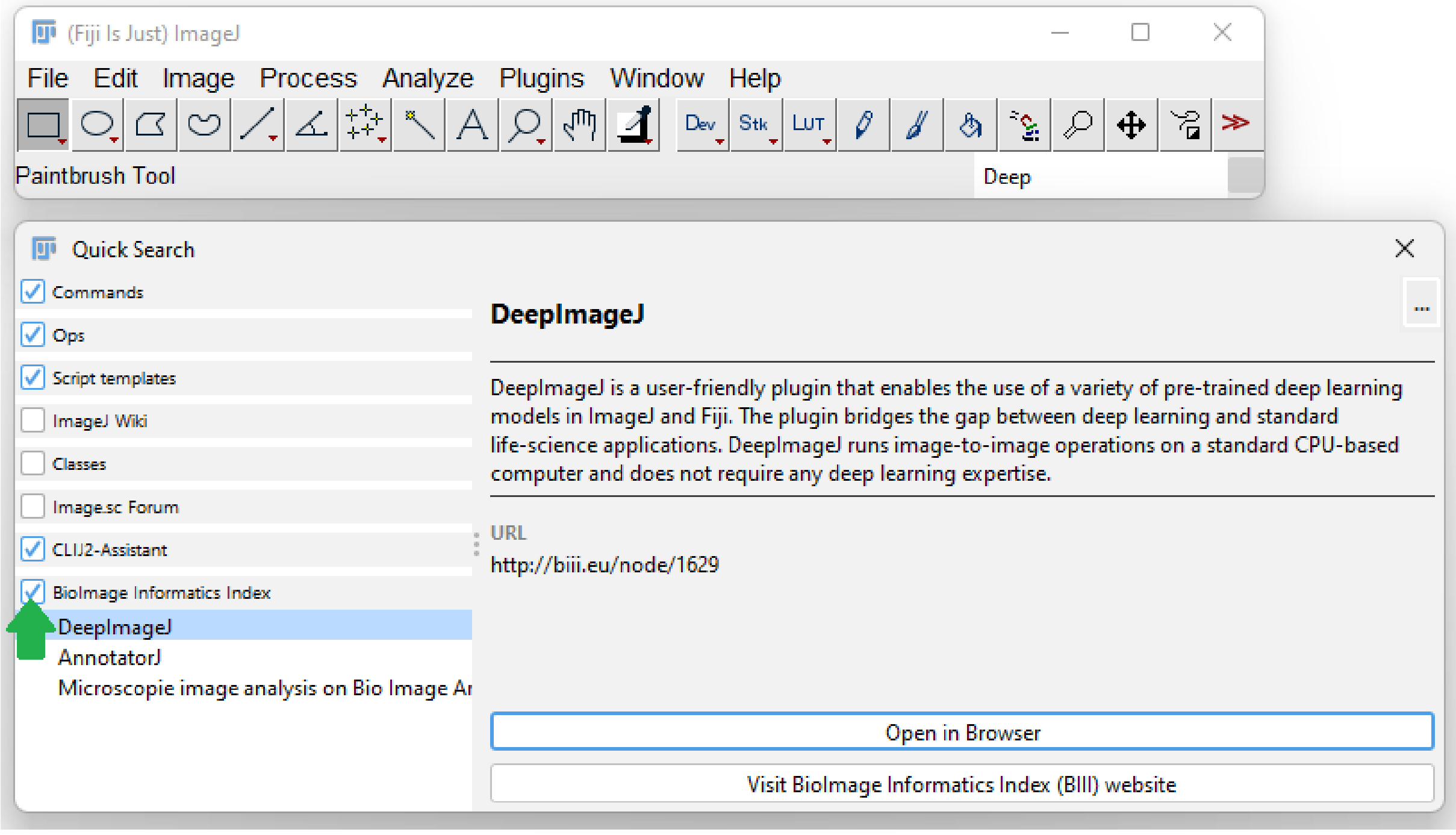

*Figure S5: The NEUBIAS search bar*

## 2.2 Data mining within BIII content

As described in Supplementary note 2, Bise Core Ontology has been designed to model and better share the content of the biii.eu bioimaging resources repository. EDAM Bioimaging aims at capturing domain-specific knowledge related to bioimaging data analysis in terms of topics and operations. We briefly illustrate sample queries benefiting from knowledge captured in these two ontologies through a set of Jupyter notebooks, which can be launched for example using Binder as demonstrated in the github page for interactive edition: [https://github.com/NEUBIAS/bise-core-ontology](https://github.com/NEUBIAS/bise-core-ontology)

These notebooks commonly first load an RDF dataset which is automatically updated once a week. They are based on SPARQL[4] language requests on the graph constructed from the RDF dataset, and python.

### 2.2.1 Curation oriented queries

An example notebook provides some curation oriented queries:

[bise-core-ontology/quality-curation-queries.ipynb at master · NEUBIAS/bise-core-ontology (github.com)]()

These queries filter Biii entries that do not achieve yet a comprehensive enough description, and can serve as an indicator of progress for Taggathons.

### 2.2.2 Inferring software author communities based on shared interests

Here we demonstrate how subgraphs can be constructed to assemble a network of bioimaging tool authors, based on shared interest. Here we make the assumption that two authors are connected if they developed different tools annotated with the same EDAM operation (Figure S6).

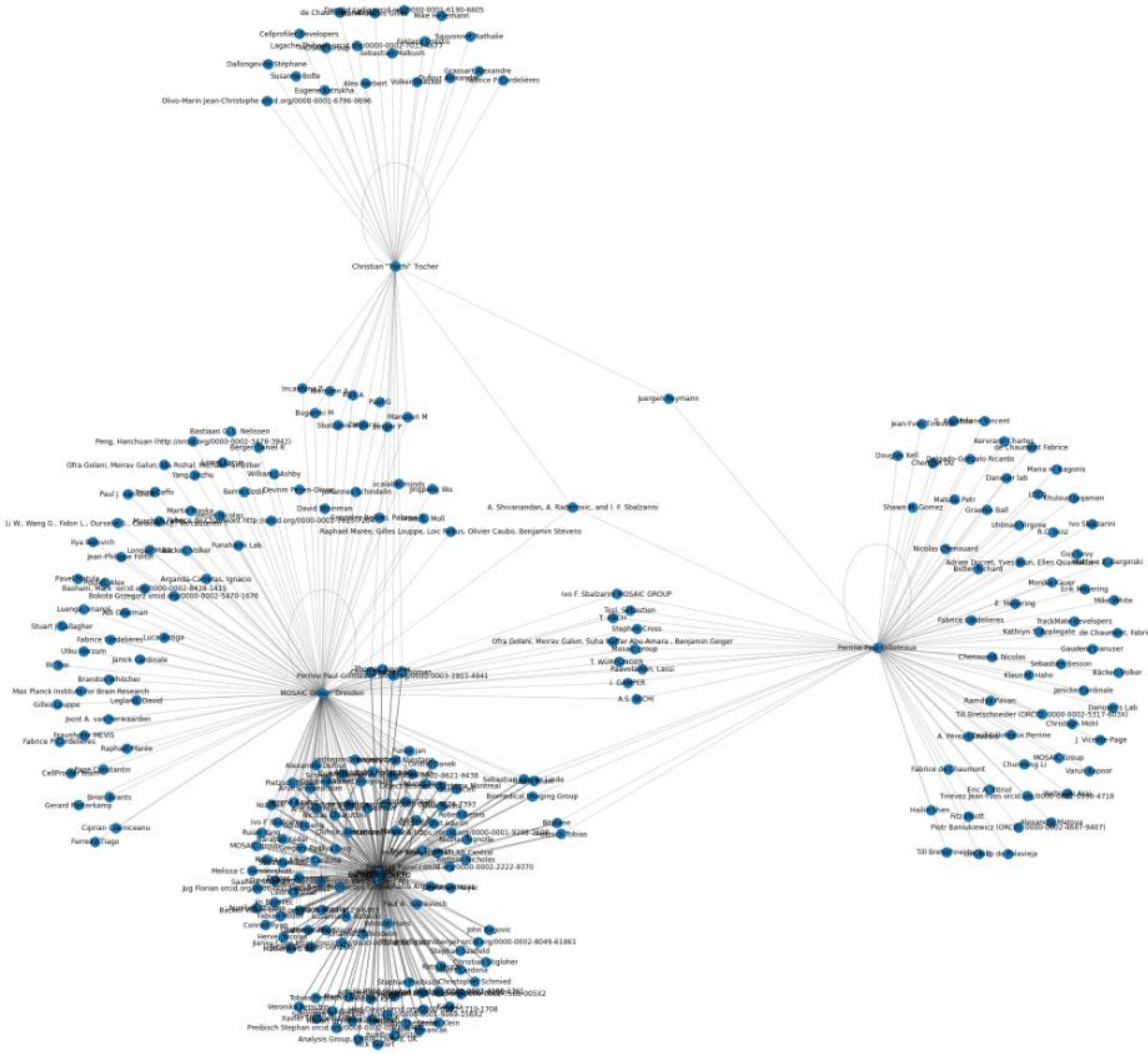

***Figure S6:*** *Screenshot of the graph generated by the notebook, showing connection between authors based on EDAM operations.*